\documentclass[runningheads]{llncs}
\usepackage{graphicx}
\usepackage{booktabs}

\begin{document}
\title{Automated ensemble method for pediatric brain tumor segmentation}
%
%
\author{Shashidhar Reddy Javaji\inst{1*} \and
Advait Gosai\inst{1*} \and
Sovesh Mohapatra\inst{1,2*} \and
Gottfried Schlaug\inst{2,3,4}
}
\authorrunning{S.R. Javaji et al.}
%
\institute{Manning College of Information and Computer Sciences, University of Massachusetts Amherst, MA, 01002 \and
Institute for Applied Life Sciences, University of Massachusetts Amherst, MA, 01003 \and
Department of Biomedical Engineering, University of Massachusetts Amherst, MA, 01002\and
Department of Neurology, Baystate Medical Center, and UMass Chan Medical School - Baystate Campus, Springfield, MA 01199\\
\email{gschlaug@umass.edu}}

\maketitle              

\small{* contributed equally} \\
\begin{abstract}
Brain tumors remain a critical global health challenge, necessitating advancements in diagnostic techniques and treatment methodologies. A tumor or its recurrence often needs to be identified in imaging studies and differentiated from normal brain tissue. In response to the growing need for age-specific segmentation models, particularly for pediatric patients, this study explores the deployment of deep learning techniques using magnetic resonance imaging (MRI) modalities. By introducing a novel ensemble approach using ONet and modified versions of UNet, coupled with innovative loss functions, this study achieves a precise segmentation model for the BraTS-PEDs 2023 Challenge. Data augmentation, including both single and composite transformations, ensures model robustness and accuracy across different scanning protocols. The ensemble strategy, integrating the ONet and UNet models, shows greater effectiveness in capturing specific features and modeling diverse aspects of the MRI images which result in lesion wise Dice scores of 0.52, 0.72 and 0.78 on unseen validation data and scores of 0.55, 0.70, 0.79 on final testing data for the "enhancing tumor", "tumor core" and "whole tumor" labels respectively. Visual comparisons further confirm the superiority of the ensemble method in accurate tumor region coverage. The results indicate that this advanced ensemble approach, building upon the unique strengths of individual models, offers promising prospects for enhanced diagnostic accuracy and effective treatment planning and monitoring for brain tumors in pediatric brains. 

\end{abstract}
\section{Introduction}
Brain tumor has been a major global health challenge, impacting not only adults but also children and adolescents \cite{1}. In 2022, the United States alone reported an estimated 40,594 individuals, from infancy to 19 years of age, diagnosed with primary brain or other central nervous system (CNS) tumors. Among these, pilocytic astrocytoma was the predominant histopathologic group, accounting for 8,264 cases. Given that survival rates following a diagnosis are particularly low among infants, there is a huge need to develop age-specific segmentation models \cite{2}. These models could enable precise and automated detection of tumor regions within pediatric brains, thereby facilitating more efficient and fast diagnosis and treatment. 

With the increasing use of deep learning techniques in conjunction with different modalities of MRI, different models, especially the U-shaped architectures, have demonstrated precise and accurate performance across various medical image segmentation tasks \cite{3,4,5}. Various models have been employed for whole brain segmentation, yet certain architectural designs have been observed to perform better in segmenting specific regions within the brain \cite{6}. This phenomenon can be attributed to the intricate relationship between the complexities inherent in both the architecture and the brain's structure. Just as different modalities provide complementary information about anomalous regions in the brain, different architectures may be more adept at handling particular areas of the brain\cite{7}.

In this paper, we are focused on using the CBTN-CONNECT-DIPGr-ASNR-MICCAI BraTS-PEDs 2023 Challenge data which constitutes four different modalities (native T1, post-contrast T1-weighted (T1Gd), T2-weighted (T2), and T2 Fluid Attenuated Inversion Recovery (T2-FLAIR)) \cite{8}. We introduce a novel ensemble approach using ONet and various modified versions of UNet along with modified loss functions which yielded a more precise segmentation for the pediatric tumors.

\section{Methodology}

\begin{figure}[h!]
\centering
\includegraphics[width=0.8\textwidth]{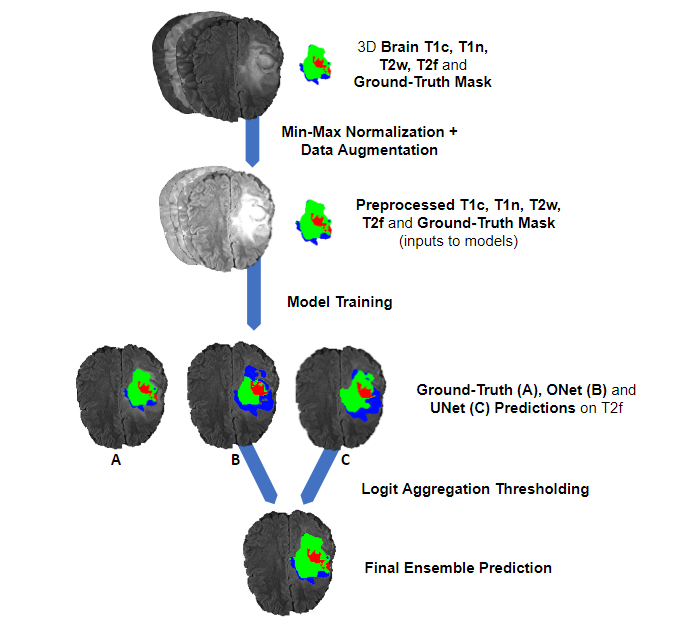}
\caption{Comprehensive workflow visualizing the end-to-end process with model training and ensemble approaches.} \label{fig1}
\end{figure}

\subsection{Data Augmentation}
Due to the inherent variability and noise that can occur during acquisition using different scanners, the integration of various physics-inspired augmentation techniques is essential. By populating the training data with these techniques, the model becomes robust and accurate across diverse scanning protocols.\cite{9}

In this study, we used two different approaches of augmentation: single and composite transformation. For the single transformation, we used techniques like flip, affine transformation, elastic deformation, noise, rescale Intensity, and random bias field. Additionally, in composite transformation, a combination of some or all of the aforementioned techniques was applied to further enrich the dataset.\cite{10,11,12,13}
\subsection{Model Architectures}

\subsubsection{UNet3D Family}
In this work, we implemented eight unique variations using two fundamental base architectures, each paired with different loss functions (further explained in Section \ref{lossf}) and tailored hyperparameters. Figure \ref{unetfig} illustrates the architecture for one of the variants of the UNet3D configurations. 

\begin{figure}[h!]
\includegraphics[width=\textwidth]{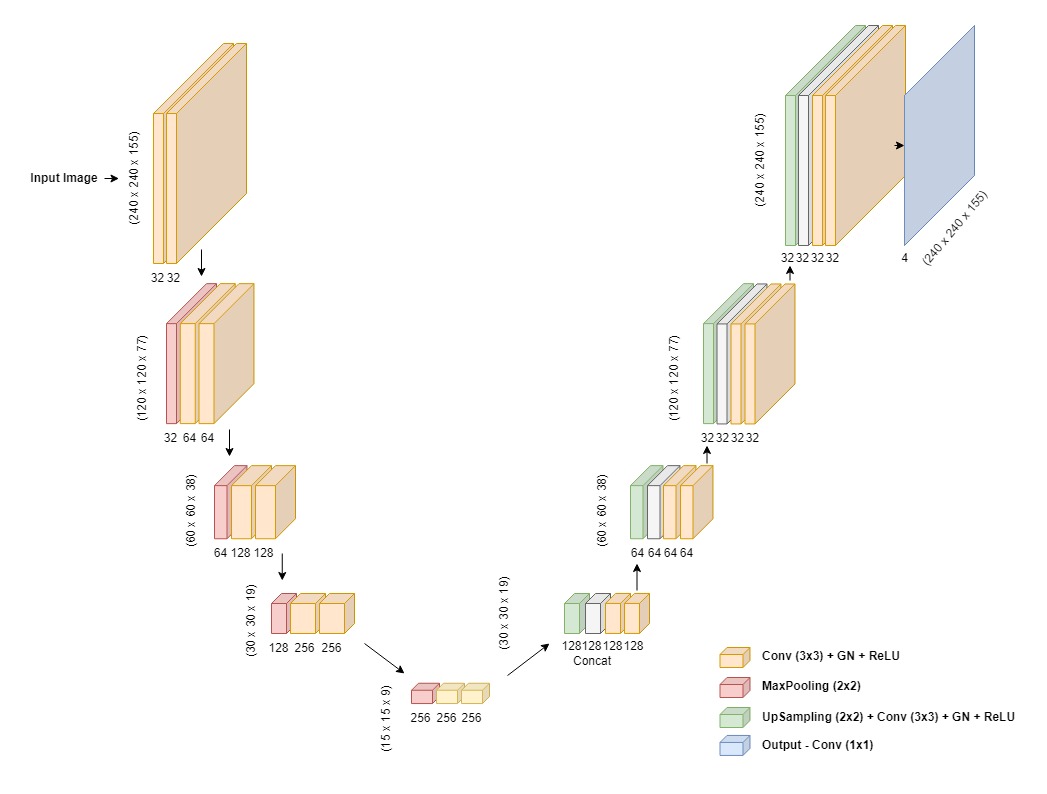}
\caption{Core UNet3D Architecture.}
\label{unetfig}
\end{figure}

The UNet variations are as follows:

\begin{itemize}
    \item \textbf{3D UNet}: This is the standard configuration utilizing an input channel size of 4, output class size of 3, and a channel size of 32.
    \item \textbf{3D UNet GELU}: A variant of the standard 3D UNet where the ReLU activation function is substituted with GELU (Gaussian Error Linear Unit) to potentially improve non-linear learning capabilities.
    \item \textbf{3D UNet SingleConv}: This version modifies the typical double convolution process in each upscaling and downscaling step, replacing it with a single convolution, which might affect the model's ability to capture complex features.
    \item \textbf{3D UNet Attention}: Incorporates an attention mechanism into the 3D UNet structure, aiming to enhance the model's focus on relevant features for improved segmentation.
    \item \textbf{3D UNet Dropout}: Introduces dropout layers into the 3D UNet architecture to prevent overfitting and promote generalization.
\end{itemize}

\subsubsection{ONet3D Family}

We also explore variations within the ONet3D family, which differs from the UNet3D by concatenating the encoder-decoder sections before the output convolution layer:

\begin{itemize}
    \item \textbf{3D ONet SingleConv Kernel\_1}: Utilizes a single convolution with a kernel size of 1 in the ONet3D architecture.
    \item \textbf{3D ONet SingleConv Kernel\_5}: Adopts a single convolution with a kernel size of 5, offering a broader receptive field compared to Kernel\_1.
    \item \textbf{3D ONet DoubleConv Kernel\_1}: Features double convolution operations with a kernel size of 1, potentially enhancing the model's ability to extract fine-grained features.
\end{itemize}

\subsection{Loss Functions}\label{lossf}
In our segmentation framework, we explored different loss functions to optimize the model performance. Specifically, we employed two distinct loss combinations: Binary Cross-Entropy (BCE)\cite{14} plus Dice Loss, and Generalized Dice Loss\cite{15} plus BCE. The choice of these loss functions is motivated by their abilities to tackle the class imbalance problem often encountered in medical image segmentation.

\subsubsection{Binary Cross-Entropy combined with Dice Loss}
The Binary Cross-Entropy (BCE) loss is commonly used for binary classification tasks and computes the cross-entropy between the true labels and predicted probabilities. To enhance the model's sensitivity to the region of interest, we combined BCE with Dice Loss. The Dice Loss computes the overlap between the predicted segmentation and the ground truth, providing a spatially aware metric that is sensitive to the shape of the segmented structures.

The BCE loss is defined as:

\begin{equation}
BCE(y,\hat{y}) = -\frac{1}{N} \sum_{i=1}^{N} \left( y_i \log(\hat{y}_i) + (1 - y_i) \log(1 - \hat{y}_i) \right)
\end{equation}

and the Dice Loss is defined as:

\begin{equation}
Dice(y,\hat{y}) = 1 - \frac{2 \sum_{i=1}^{N} y_i \hat{y}_i + \varepsilon}{\sum_{i=1}^{N} y_i + \sum_{i=1}^{N} \hat{y}_i + \varepsilon}
\end{equation}

where \( \varepsilon \) is a small constant to avoid division by zero. The combined loss function is:

\begin{equation}
Loss(y,\hat{y}) = \alpha \cdot BCE(y,\hat{y}) + \beta \cdot Dice(y,\hat{y})
\end{equation}

where \( \alpha \) and \( \beta \) are weights that balance the contribution of each term.

\subsubsection{Generalized Dice Loss combined with BCE}
Recognizing the limitations of standard Dice Loss in handling cases with varying object sizes, we also experimented with Generalized Dice Loss (GDL). GDL extends the traditional Dice Loss by incorporating class-wise weights, thus accommodating the imbalanced class distribution. This loss function is especially well-suited to medical image segmentation where certain classes or labels may be underrepresented.

The GDL is defined as:

\begin{equation}
GDL(y,\hat{y}) = 1 - \frac{2 \sum_{c=1}^{C} w_c \sum_{i=1}^{N} y_{ci} \hat{y}_{ci} + \varepsilon}{\sum_{c=1}^{C} w_c (\sum_{i=1}^{N} y_{ci} + \sum_{i=1}^{N} \hat{y}_{ci}) + \varepsilon}
\end{equation}

where \( C \) is the number of classes, \( w_c \) is the weight for class \( c \), defined as:

\begin{equation}
w_c = \frac{1}{(\sum_{i=1}^{N} y_{ci})^2}
\end{equation}

Here, \( \varepsilon \) is again a small constant to avoid division by zero.

\subsection{Ensemble Strategy}

The ensemble strategy integrates the predictive capabilities of ONet and UNet models, capitalizing on their complementary strengths. Initial predictions from the models are combined through a summation of logits, followed by thresholding. A majority vote across the models for each voxel finalizes the ensemble prediction, aiming to capture specific features and model diverse aspects of the MRI images.

\subsection{Post Processing}

Post-processing techniques are applied to the combined predictions to refine segmentation quality:

\begin{itemize}
    \item Size Filtering Based on Voxel Volumes: This technique removes small isolated regions, eliminating noise.
    \item Morphological Reconstruction: A more advanced method for interpolation of voxels violating constraints and smoothing boundaries for all labels. 
\end{itemize}

\section{Results}

\subsection{Evaluation Metrics}

The BraTS challenge evaluates submitted models using two primary metrics: the lesion-wise (LW) Dice score and the 95th percentile lesion-wise Hausdorff distance (HD95). These metrics are used to assess segmentations across three distinct tumor sub-regions: the whole tumor (WT), tumor core (TC) , and enhancing tumor (ET).

The Lesion-wise Dice Score and 95th Percentile Lesion-wise Hausdorff Distance (HD95) are key metrics for evaluating segmentation models in medical imaging. The Dice score, ranging from 0 (no overlap) to 1 (perfect overlap), measures the accuracy of predicted segmentations against the true segmentations on a lesion-by-lesion basis, penalizing False Positives and False Negatives by assigning a 0 score. The HD95 metric quantifies the maximum deviation between predicted and actual segmentations for each lesion, with False Positives and False Negatives receiving a fixed penalty value of 374. Mean scores for both metrics are calculated across case IDs. These evaluations offer a detailed insight into the model's segmentation performance, highlighting areas for potential enhancement

\subsection{Quantitative Performance}

\begin{table}[h!]
\centering
\caption{Comparison of evaluation metrics for Enhancing Tumor on validation data}
\begin{tabular}{l c c c c}
\toprule
Models & LW Dice $\uparrow$ & Dice $\uparrow$ & LW HD95 $\downarrow$ & HD95 $\downarrow$\\
\midrule
UNet3D          & 0.43 & 0.44 & 168.68 & 114.03\\
ONet3D          & 0.52 & 0.48 & 131.28 & 121.74\\
Ensemble (BCE, Dice) & 0.38 & 0.38 & 186.73 & 141.84\\
Ensemble (BCE, GD)   & \textbf{0.52} & \textbf{0.49} & \textbf{127.11} & \textbf{105.46} \\
\bottomrule
\end{tabular}
\end{table}

\begin{table}[h!]
\centering
\caption{Comparison of evaluation metrics for Tumor Core on validation data}
\begin{tabular}{l c c c c}
\toprule
Models & LW Dice $\uparrow$ & Dice $\uparrow$ & LW HD95 $\downarrow$ & HD95 $\downarrow$\\
\midrule
UNet3D & 0.70 & 0.75 & 44.68 & 12.19\\
ONet3D & 0.71 & 0.74 & 33.26 & 19.82\\
Ensemble (BCE, Dice) & 0.64 & 0.73 & 60.00 & 17.68\\
Ensemble (BCE, GD) & \textbf{0.72} & \textbf{0.74} & \textbf{23.84} & \textbf{11.48} \\
\bottomrule
\end{tabular}
\end{table}

\begin{table}[h!]
\centering
\caption{Comparison of evaluation metrics for Whole Tumor on validation data}
\begin{tabular}{l c c c c}
\toprule
Models & LW Dice $\uparrow$ & Dice $\uparrow$ & LW HD95 $\downarrow$ & HD95 $\downarrow$\\
\midrule
UNet3D & 0.75 & 0.82 & 45.51 & 11.85\\
ONet3D & 0.76 & 0.80 & 32.88 & 19.31\\
Ensemble (BCE, Dice) & 0.70 & 0.82 & 61.74 & 15.97\\
Ensemble (BCE, GD) & \textbf{0.78} & \textbf{0.82} & \textbf{25.88} & \textbf{10.89} \\
\bottomrule
\end{tabular}
\end{table}

In our assessment of various models, we submitted the results to the synapse portal for testing the results on the validation. The findings revealed that the ensemble training approach demonstrates better effectiveness in comparison to the single model training approach, across almost  all of the validation cases.

Furthermore, an individual evaluation of predictions indicated that the single-model training approach of the ONet3D model matches the ensemble training approach for lesion-wise dice as outlined in Table 1. This indicates that there can be an ensemble model which might be robust and be generalizable for a larger dataset. However for specific types of segmentation and evaluation metrics, the approaches might differ. 

\begin{table}[h!]
\centering
\caption{Evaluation metrics for sub-regions on \textbf{Final Test Data}}
\begin{tabular}{l c c c c}
\toprule
Sub-region & LW Dice $\uparrow$ & Dice $\uparrow$ & LW HD95 $\downarrow$ & HD95 $\downarrow$\\
\midrule
ET & \textbf{0.55} & \textbf{65.23} & \textbf{0.55} & \textbf{0.99} \\
TC & \textbf{0.71} & \textbf{31.61} & \textbf{0.61} & \textbf{0.99} \\
WT & \textbf{0.79} & \textbf{22.36} & \textbf{0.70} & \textbf{0.99} \\

\bottomrule
\end{tabular}
\end{table}

In the evaluation of our brain tumor segmentation model on the BraTS test set, the results demonstrate varying degrees of performance across different tumor sub-regions, indicating the model's proficiency in distinguishing between tumor core (TC), enhancing tumor (ET), and whole tumor (WT) regions. Specifically, the model achieved Dice scores of 0.5519, 0.7054, and 0.7938 for ET, TC, and WT, respectively, suggesting a higher accuracy in segmenting the whole tumor region compared to the more challenging enhancing tumor and tumor core regions. Similarly, the Hausdorff distance (95th percentile) metrics, which assess the model's precision in outlining tumor boundaries, show values of 65.23, 31.61, and 22.36 for ET, TC, and WT, respectively, indicating the model's increased boundary precision in WT segmentation. The sensitivity scores—0.5544 for ET, 0.6096 for TC, and 0.7010 for WT—further highlight the model's ability to correctly identify tumor pixels, with WT regions being most accurately detected. In contrast, the specificity scores, which are consistently high across all regions (0.9997 for ET, 0.9998 for TC, and 0.9999 for WT), underscore the model's effectiveness in correctly classifying non-tumor pixels. These results collectively underscore the model's robustness in brain tumor segmentation, with notable strengths in whole tumor delineation and high specificity across tumor sub-regions. In the end it's evident from the validation and test results that the model generalizes well.

\subsection{Visual Comparison}

Figure 3 illustrates a side-by-side comparison of prediction results derived from two separate methodologies for modeling: individual model training and the ensemble approach. 

The comparison reveals a noticeable difference between the two approaches. The predictions stemming from the single model training method, as evidenced by UNet3D and ONet3D, demonstrate to capture less core tumor region and region affected by the tumor (enhancing tumor). This limitation is contrasted by the predictions generated through the ensemble training method.

\begin{figure}[h!]
\centering
\includegraphics[width=0.85\textwidth]{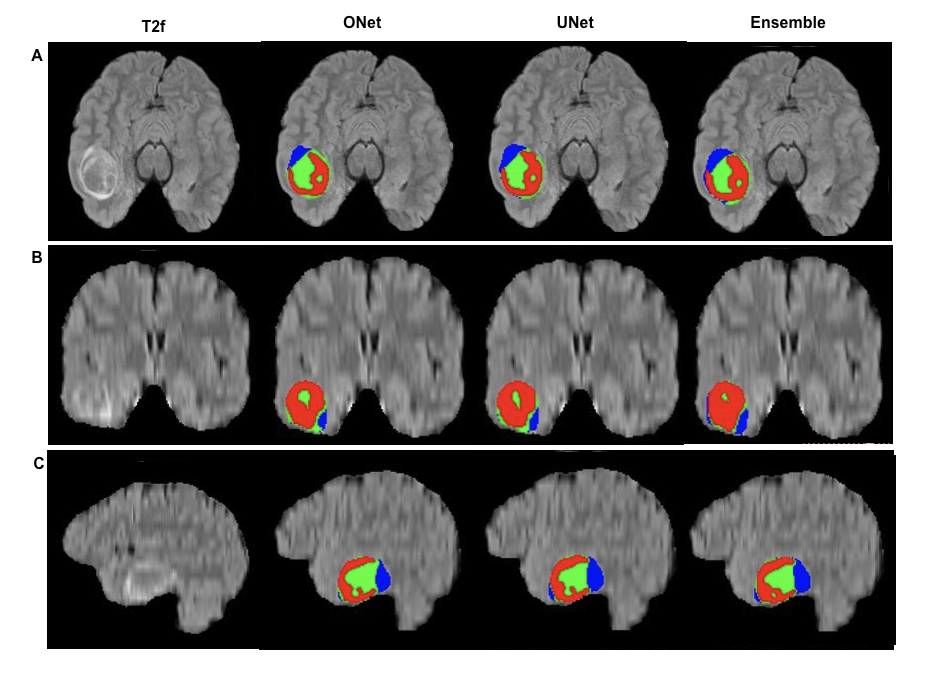}
\caption{A, B, and C depict three slice planes (sagittal, coronal, axial) from the same CaseID, showcasing predictions generated by three different models—UNet3D, ONet3D, and the ensemble model respectively.}
\label{unetfig}
\end{figure}

The ensemble approach adds the strengths of multiple individual models, potentially leading to more accurate and robust predictions. In the context of the figure, it becomes evident that employing the ensemble method results in more effective coverage of tumor regions. This might suggest that the ensemble training not only compensates for the limitations observed in the single model training but possibly enhances the overall predictive capability.

\section{Discussion and Conclusion}

The integration of ONet and UNet models through an ensemble technique, together with the novel proposed post-processing strategy, creates an advanced approach to medical image segmentation. This method builds upon the unique strengths of both ONet and UNet models, combining them in an interdependent manner that leverages their individual capabilities. By doing so, the approach not only amplifies the robustness of the segmentation but also adds a level of precision that might be unattainable with a single-model training strategy. The clinical significance of this methodology lies in its potential to offer enhanced diagnostic accuracy and effective treatment planning. 

%
%
%
%

\end{document}